\documentclass[12pt,preprint]{aastex}

\usepackage{multirow}

\newcommand{\Zmax}{Z_{\rm max}~}

\shorttitle{Moving Group Catalog} \shortauthors{Zhao et al.}

\begin{document}


\title{A Catalog of Moving Group Candidates in The Solar
   Neighborhood}


\author{Jingkun Zhao, Gang Zhao, Yuqin Chen}

\altaffiltext{}{National Astronomical Observatories, Chinese
Academy of Sciences, Beijing, 100012, China; gzhao@bao.ac.cn.}


\begin{abstract}
Based on kernel estimator and wavelet technique, we have
identified 22 moving group candidates in the solar neighborhood
from a sample which includes around 14000 dwarfs and 6000 giants.
Six of them were previously known as the Hercules stream, the
Sirus-UMa stream, the Hyades stream, the Caster group, the
Pleiades stream and the IC 2391; five of them have also been
reported by other authors. Eleven moving group candidates, not
previously reported in the literature, show prominent structures
in dwarf or giant samples. A catalog of moving group candidates in
the solar neighborhood is presented in this work.
\end{abstract}




\section{Introduction}           
\label{sect:intro}

Since the first discovery of the two clearest examples of moving
groups in the solar neighborhood (the Hyades and the Ursa) by
Proctor (1869), the studies of moving group have been paid
attention for a long time. There are two main techniques to detect
moving groups. One is the convergent point (Brown 1950, Jones
1971) technique for proper motion data; the other is searching
kinematic structure in velocity space. Most works adopted the
second technique. For example, G$\acute{o}mez$ et al. (1990) used
the SEM algorithm to decompose the sample into the sum of
tridimensional gaussians in the (U,V,W) velocity, and concluded
that the observed distribution of the residual velocity can be
explained as the sum of four independent distributions which were
related to several open clusters in the solar neighborhood. Chen
et al. (1997) developed an algorithm using a non-parametric kernel
estimator to describe the the stellar distribution in a
4-dimensional space ( velocities and age) and four moving groups
near the Sun (Pleiades, Sirius, Hyades, IC2391) have been
identified without assuming any prior knowledge of moving groups,
neither the velocity distribution nor other physical properties.

After the release of the Hipparcos Catalogue (ESA 1997), the
research of moving groups in the solar neighborhood has made great
progress since accurate parallaxes and proper motions for a large
number of stars are available. (Skuljan et al. 1999) use the
wavelet transform technique to analyze the distribution derived by
an adaptive kernel method and also find several moving groups.
Recent works by Famaey et al. (2005, 2007) have identified five
moving groups based on 6000 giants and investigated the mass
distribution of the Hyades stream based on the Geneva-Copenhagen
survey of 14000 dwarfs from Nordstr\"{o}m et al. (2004). Klement
et al. (2008) identified at least four `phase space overdensities'
of stars on very similar orbits in the solar neighborhood using
the $1^{st}$ RAVE public data release. Different authors use
different techniques and different samples, but they all report
the presence of moving groups in the solar neighborhood.

With the newly-developed wavelet transform technique by Skuljan et
al. (1999), we attempt to identify the stellar moving groups in
the solar neighborhood by combining both the dwarf samples from
Nordstr\"{o}m et al. (2004) and the giant samples from Famaey et
al. (2005). The main goals are to investigate whether the
locations in velocity space of those groups derived from different
samples are consistent as well as finding new streams based on
large samples of stars and new method.


\section{The method to identify moving group }
Since the sample is in the solar neighborhood, the position of the
stars does not provide any discriminant information for the
detection of moving groups, so the detection in our paper mainly
depends on the two components of the stellar velocity-the U
component pointing towards the galactic, V component towards the
direction of the galactic rotation. First, the probability density
function f(U,V) is estimate by a kernel function, and next the
wavelet transform will be carried out, so we can finally recognize
the moving groups on the basis of the analysis of wavelet
coefficients.
\subsection{Density distribution}
We use the kernel function to decide the probability density at
any given point. The type of kernel function used in our method is
a radial basis function (Equation 1). h is a smoothing parameter,
($\Sigma$) is the covariance matrix. The probability density of
the (U,V) panel is then derived by Equation 2. n is the number of
stars in our sample.
\begin{eqnarray}
K(x,x_i)&=&\exp(-\frac{1}{2*h^2}(x-x_i)^{T}(\Sigma)^{-1}(x-x_i))
~~~h>0
\end{eqnarray}

\begin{eqnarray}
\rho(x)&=&\frac{1}{nh^22\pi|\Sigma|^{1/2}}\sum_{i=1}^{n}K(x,x_i)
\end{eqnarray}
\subsection{Wavelet transform}
After deriving the density of any given (U,V) point, in order to
achieve the (U,V) center and the dispersion of the possible moving
groups, we use two-dimension wavelet transform technique to
analyze them (Skuljan et al. 1999). Wavelet analysis (Daubechies
1988; Chui 1992; Ruskai et al. 1992) is a fantastic tool and is
becoming more and more popular in astronomical researches.

   The wavelet transform provides an easily interpretable visual
representation of $\rho(x)$. Moreover the continuous wavelet
transform can be used in singularity detection. In this paper, we
use it to find some structures working at different scales.

   To process a wavelet transform of a function f(x,y), we must
define an analyzing wavelet $\psi(x/\sigma,y/\sigma)$ called
mother wavelet in advance. $\sigma$ is scale variable. The wavelet
coefficient of the point $(\mu,\nu)$ then can be derived by
Equation 3.

\begin{eqnarray}
\omega(\mu,\nu)&=&\int_{-\infty}^{\infty}\int_{-\infty}^{\infty}f(x,y)\psi(\frac{x-\mu}{\sigma},\frac{y-\nu}{\sigma})dxdy
\end{eqnarray}
The actual choice of the analyzing wavelet $\psi$ depends on the
particular application. A mother wavelet named Mexican hat
(Skuljan et al. 1999) is the second derivative of a Gaussian and
generally gets good results when applied to find singularities. So
when we intend to search for certain groups from a given data
distribution, a two-dimensional Mexican hat is selected as the
mother wavelet (Equation 4).
\begin{eqnarray}
\psi(x,y,a)&=&(2-\frac{x^2+y^2}{a^2})e^{-(x^2+y^2)/2a^2}
\end{eqnarray}
 The main characteristic of the function $\psi$ is that the total
volume is equal to zero, which is what enables us to detect any
over-densities in our data distribution. The wavelet coefficients
will be all zero if the analyzed distribution is uniform; but if
there is any significant `bump' in the distribution, the wavelet
transform will give a positive value at that point.

\section{Results and Discussions}

 \label{sect:RaV}

Two data sets are selected as our samples. One is the 14000 dwarf
(Nordstr\"{o}m et al. 2004, hereafter Nord04 ); the other is 5311
K and 719 M giant (Famaey et al. 2005). The information of dwarf
samples came from the Geneva-Copenhagen survey, in which the ages,
metallicities and kinematic properties are provided, while that of
the giant sample was obtained from CORAVEL/Hipparcos/Tycho-2.
First the probability density was derived by the method (Sec 2.1).
For equation 1, the smoothing parameter h should be carefully
selected because small values of h emphasize noisy structure while
large h value will smooth out all the details of the distribution.
Our selection of the value h adopts the method of Asiain et al
(1999). The value of h is set to be 0.145 for the dwarf sample and
0.117 for the giant sample.  The density distribution in the (U,V)
plane has some irregularities and can not be described by a unique
velocity ellipsoid, and some `bumps' can be clearly recognized.
Famaey et al.(2008)derived the threshold of noise wavelet
coefficients to have a value of $10^{-4}$. Thus, the coefficients
smaller than $10^{-4}$ in the present work will be rejected. Fig.1
shows the contour map of the positive wavelet coefficients
obtained in the (U,V) space from $a$ = 4km$s^{-1}$.

\subsection{Monte Carlo simulation}
   After obtaining the probability density of the samples and the
wavelet coefficient contour map, the results show several
features. However, are these features real or are they caused by
some noise? There are several factors that will affect the
observed number of stars in each bin: such as the measurement
error, statistical fluctuations related to the finite sample, etc.
   We can expect that the number of the observed stars in each bin
will subject Poisson distribution with an average of $\bar{N}$
which can be derived by the probability density function
multiplied with factor nS, where n is the total number of stars
and S is area covered by a square bin.
   To investigate how probable are the features showed in  Fig.1, we generate
a large number (N=2000) of Poisson random copies. For each copy,
we get a wavelet positive coefficient map (after rejecting those
smaller than $10^{-4}$). Fig.2 is a example of one copy. The
probability of those features is shown in Table 1. Typically, only
those features with 90 percent or better probability are
considered to be real.

\subsection{The detected moving groups}

From Fig.1, it is clear that stars are clumped at different
locations in the (U,V) space. Table 1 shows the central (U,V) of
22 possible moving groups detected by either the dwarf sample or
the giant sample by using the wavelet technique. We analyzed the
dwarf sample and the giant star sample separately because the star
numbers in the solar neighborhood of dwarf and giant stars are
different due to different lifetimes with more dwarf stars than
giant stars. In view of this, merging the two types of stars into
one sample will reduce the statistically grouped structure in the
(U,V) space. In some cases, the giant sample can give better
structure than the dwarf sample as we will show later (for group
7). It seems that a more reasonable way is the comparison of these
grouped structures detected from both the dwarf and giant sample
and check if they can give consistent results. In this sense, we
have found that the central (U,V) values of these groups between
the two samples are quite consistent from 12 groups.

Among these groups, groups 1 to 5 are quite strong  and they have
been widely reported to be moving groups in the literature. As we
can compare our results with previous works by Eggen (1991, 1992a,
b, c) based on the FK5 catalogue and Dehnen (1998), we found good
agreement for the Pleiades (group 1), Hyades (group 2) and Sirius
(group 4). Specifically, our values are quite consistent with
those of Dehnen (1998), who gave mean motions of (-12, -22), (-40,
-20.0), (9, 3), km$s^{-1}$ for the Pleiades, Hyades and Sirius
groups, respectively, based on the sample of 14369 stars observed
by the Hipparcos satellite. Eggen (1991, 1992a, b, c) showed mean
motions of (-11.6, -20.7), (-40.0, -17.0), (14.9, 1.3) km$s^{-1}$
for the Pleiades, Hyades and Sirius are somewhat updated by using
Hipparcos parallaxes. Our values are also similar to those values
in recent works by Famaey et al. (2007) and Antoja et al (2008).
Mean motions of group 3 is also in agreement with the Hercules
stream centered at (-35, -45) found by Fux (2001). It is noticed
that the mean motions of group 3 are slightly different between
the dwarf and giant groups. Considering that the giant group shows
better structure than that of the dwarf group in Fig. 1, we
suggested that the mean motions of giant groups have more
reasonable values for group 3. Our group 5 with mean motions of
(-11, -7) in the present work is very similar to the Castor group
with (-10, -10), which is classified as a young group by Famaey et
al. (2007) with mean motion of (-10, -12).

Besides the five known moving groups, two moving groups (group 6
and group 7)  with mean motions of (38, -20) and (-57, -45) are
quite significant from both dwarf and giant samples and the
probability from our simulation is larger than $ 98\%$. For our
Group 6, Antoja et al. (2008) also suggest the structure centered
at (35, -20) to be a new group. However, since the structure is
weak in their work, they considered it as part of the elongation
of the Sirius or Coma Berenices structure. However, Group 6 is
very prominent and has a well-defined shape in the present work
both for the dwarf sample and giant sample. Moreover, their mean
motions are far away from those of Sirius and the Coma Berenices
structures without significant connections in the contour plot of
Fig. 1. Note that the giant sample of group 6 has some extension
to the south-west direction, which disappears in the simulation
contour of Fig.2 after adding noise while the center part of group
6 persists. It seems that the center part of group 6 is real while
the extension to the south-west direction may be noise. In this
sense, we suggest that group 6 is real in the giant sample, and it
is very significant in dwarf sample.

 The other five groups (8, 9, 10, 12, 14) also have some features
 in both the dwarf sample and the giant sample and the simulation shows that
four of them are significant with probability above 90\% of being
real, except for the group 10 with the probabilities of 86\% and
82\%. The rest of the groups are shown only in the dwarf sample
(15 and 16) or only in the giant sample (11, 13, 17, 18, 19, 20,
21, 22); but four groups (11, 13, 17, 21) are not very significant
with the probability lower than 90\% by simulation. Interestingly,
some of these features have been reported by previous works. For
example, our group 14 is consistent with the IC 2391 stream, which
shows a mean motion of (-20.8, -15.9) by Eggen (1992b) and of
(U=-20, V=-12) by Chen et al.(1997). Moreover, our groups 18, 20
and 22 may be the same groups as No. 7, 14 and 13 in Table 3 of
Antoja et al (2008), which summarizes the possible grouping
structures from Dehnen(1998) and Eggen's serial papers.

 A catalog of 22 possible moving group candidates are listed in
Table 1, although some of them are not so significant (10, 11, 13,
17, 21) for a statistical requirement of the probability above
90\%. Certainly, these groups with low probabilities from our
statistical analysis need further study. The metallicity is only
available in the dwarf sample, based on which we present the peak
and scatter of [Fe/H] for all the candidates in Table 1.

\subsection{The W velocity distributions of the groups}
 Although all the works to detect the moving groups do not take
into account the W velocity, this component may bring some new
information. Therefore, we have generated the (V,W) and (U,W)
contour plots with the same method. As expected, the distribution
of W velocity is limited to a narrow range from -20 km/s to 5 km/s
for both the dwarf and giant samples. Moreover, the dispersion in
the W velocity for these groups is about 10-15 km/s, which is
significantly larger than those of the U and V components of 4-6
km/s in defining these groups. Thus, the (V,W) and (U,W) contour
plots have no advantages in identifying possible moving groups.
Several groups of the 22 candidates actually belong to the same
groups in the (V,W) and (U,W) contour plots. Note that the groups
of 9 and 12 have the largest W dispersion of ~20 km/s because they
have very weak features in our Fig.1. In general, these moving
group candidates based on the (U,V) contour are considered to be
more reliable by taking into account the large scatter in the W
component distribution.

Our Group 3 and Group 7 appear to coincide with the Hercules group
identified as green lumps in Fig.1 of Famaey et al 2008 and they
think the green lumps are both the Hercules group. However, the
argument is not significant and we suggest that they may be
distinct group. Fig.3 shows the statistical result of the $\Zmax$
distribution with the cut of $\Zmax <1.0$ kpc. It seems that the
peak of Group 3 in the $\Zmax$ distribution is at 0.1-0.2 kpc
while there is a large contribution from stars with $\Zmax$ of
0.2-0.3 kpc for Group7. Moreover, from Famaey et al (2004), it is
clear that the $\rm{\sigma(U,V,W)}$ of the Hercules group is
significantly higher than those of the Hyades or Sirius, which
could indicate an overlapping of more than one distinct group. In
the present work, the $\rm{\sigma(U,V,W)}$ is nearly the same for
group 3 and group7 as well as other groups. Finally, there is also
a hint of [Fe/H] deviation with the average [Fe/H] of -0.16 dex in
group 3 versus -0.22 dex for group 7. Further study on this topic
is necessary before firm conclusions can be drawn.

\section{Conclusions}
\label{sect:Discussion}
%

Using the dwarfs and giants in the solar neighborhood, we
illustrate a detailed analysis of the UV distribution. This
analysis reveals 22 possible grouping structures identified by the
kernel estimator and wavelet technique. The location in velocity
space of 12 possible moving groups from both dwarf and giant
samples are consistent. 11 groups, including the five well-known
groups, Pleiades, Hyades, Hercules, Sirus and Castor streams, are
consistent with previous works. Eight groups, not reported by
previous works, are presented and most of them are thought to be
significant in term of statistics. In summary, a catalog of 22
moving group candidates with the centers (U,V,W), their
dispersions and mean metallicity are given.

\acknowledgments

This study is supported by the National Natural Science Foundation
of China under grants No. 10521001, 10673015, 10433010, the
National Basic Research Program of China (973 program) No.
2007CB815103/815403 and by the Ministry of Science and technology
of China under grant No. 2006AA01A120.











\clearpage

\begin{figure}
   \vspace{1mm}
\epsscale{0.8}
   \hspace{1mm}\plotone{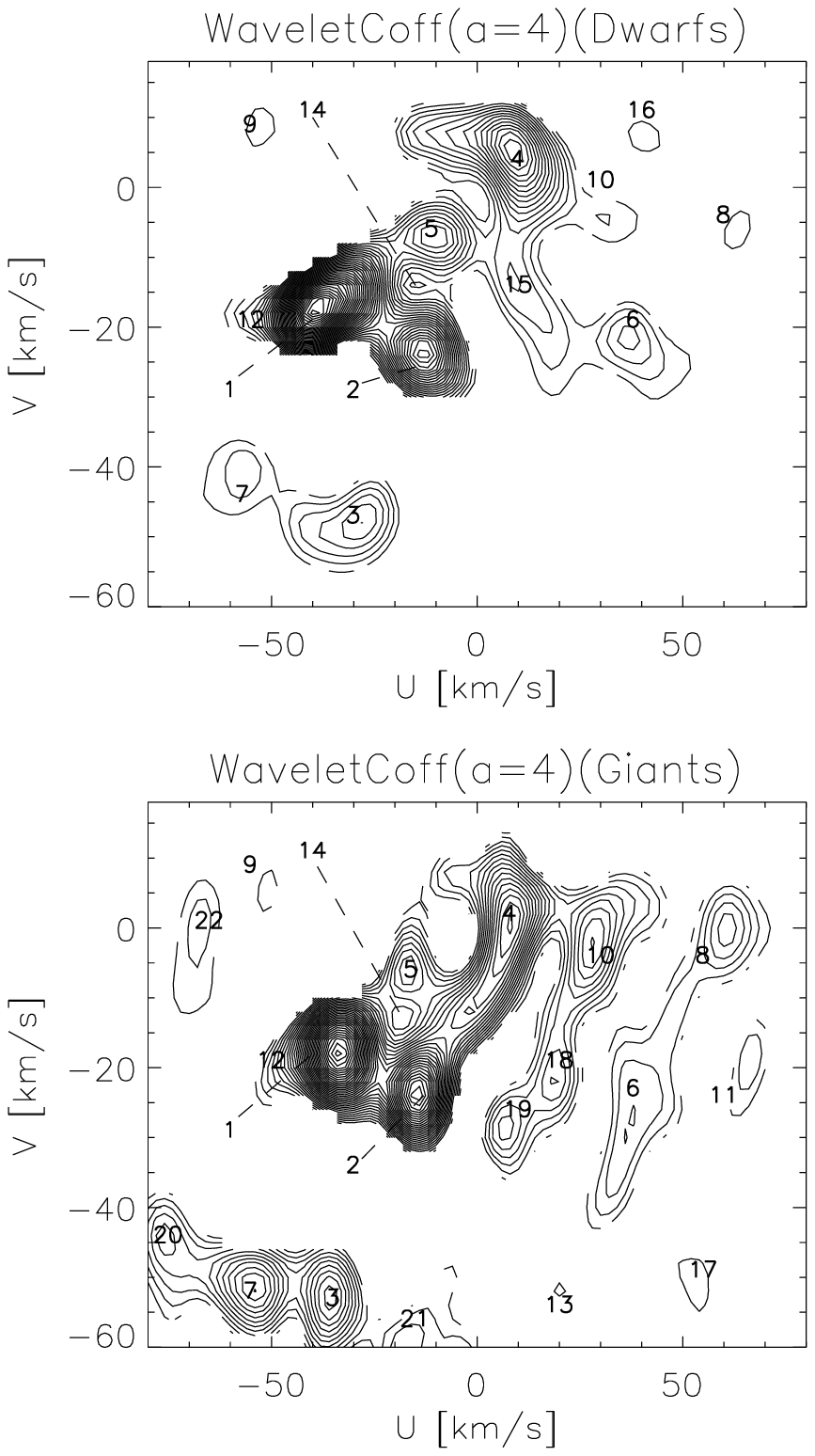}
   \caption{The contour map of the positive wavelet transform coefficient. The up is for dwarf sample and the bottom is for giant sample.
   The number is the sequence of the moving groups}
\end{figure}

\begin{figure}
   \vspace{1mm}
\epsscale{0.8}
   \hspace{1mm}\plotone{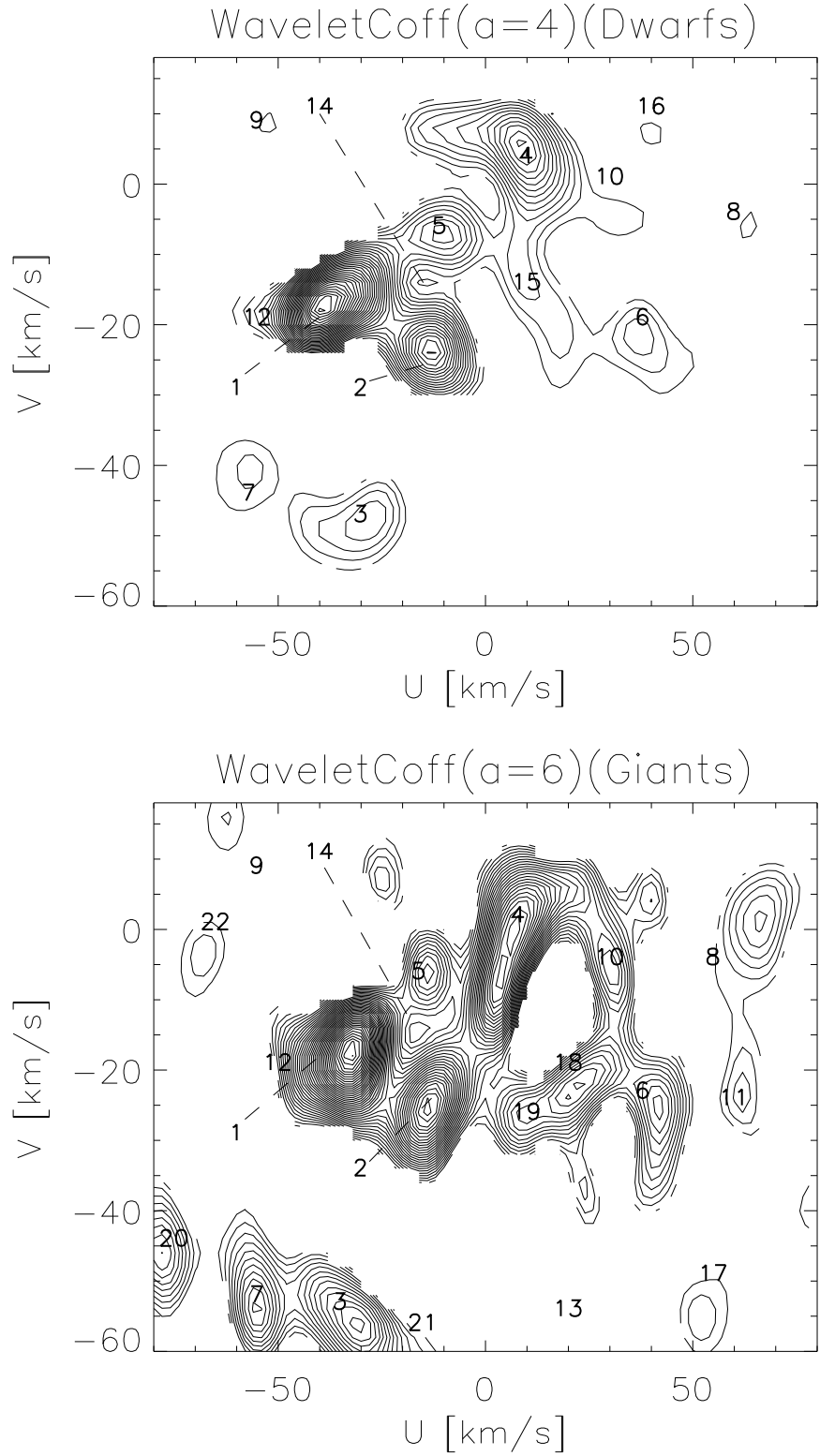}
   \caption{The contour map of positive wavelet transform
   coefficient for one copy of simulation}
\end{figure}

\clearpage
\begin{figure}
   \vspace{1mm}
\epsscale{1.0}
   \hspace{1mm}\plotone{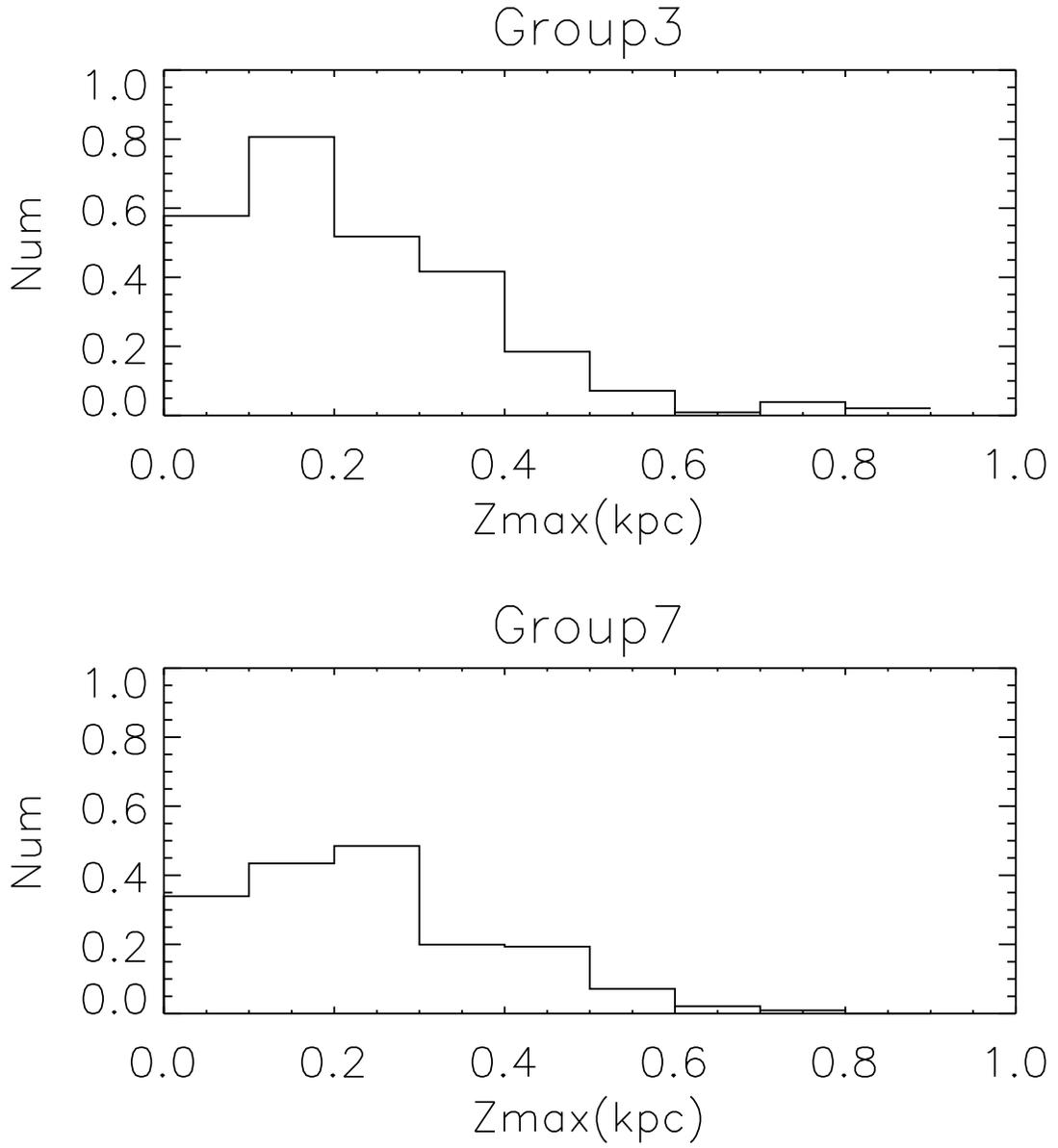}
   \caption{The Zmax distributions of Group 3 and Group 7}
\end{figure}

\begin{deluxetable}{cccccccccccccc}
 \tabletypesize{\scriptsize} \tablenum{1} \tablewidth{0pt}
  \tablecaption{  The heliocentric velocities of the moving groups detected by the dwarf sample and the giant sample.
  The column 1 is the sequence number of the groups. the column 2 and column 5 are the velocity center of the groups: column 2 is that detected by the dwarf
  sample; column 5 is that detected by the giant
  sample. Column 3 is the $\rm{\sigma(U,V,W)}$ for dwarfs and column 6 is that for giants. Column 4 is the detected probability of each moving group during the simulation for
  dwarf sample while column 7 is that for giant sample; column 8 is the average [Fe/H] of the groups; column 9 is the $\rm{\sigma[Fe/H]}$ of the groups; column 10 is the Zmax peak value
   of each group;
  the last column is the name of the corresponding moving group
identified before(A08 is from Antoja et al 2008 and T3 means Table
3 in that paper; F08 is from Famaey et al 2008 and F.3 means Fig.3
in that paper) }
\tablehead{\colhead{No}&\colhead{Dw(U,V,W)}&\colhead{$\rm{\sigma {d}(U,
V, W)}$}&\colhead{Dw(P)} &\colhead{Gi(U,V,W)}
&\colhead{$\rm{\sigma {g}(U, V,
W)}$}&\colhead{Gi(P)}&\colhead{[Fe/H]}&\colhead{$\rm{\sigma[Fe/H]}$}&\colhead{Zmax}&\colhead{Moving
Group}}
 \startdata
  \noalign{}
   1&-38,-18,-10&6,6,10&100\%&-38,-17,-11&6,6,12&100\%&-0.09&0.17&0.1-0.2&Hyades \\
   \hline
   2&-12,-23,-10&6,6,10&100\%&-15,-23,-10&6,6,12&100\%&-0.17&0.17&0.1-0.2&Pleiades \\
   \hline
   3&-32,-48,-15&5,5,12&100\%&-35,-51,-11&5,5,15&100\%&-0.16&0.20&0.1-0.2&Hercules\\
   \hline
   4&10,3,-11&6,6,10&100\%&10,3,-14&6,6,13&100\%&-0.21&0.15&0.0-0.1&Sirius-UMa  \\
   \hline
   5&-11,-7,-12&5,5,10&100\%&-13,-6,-10&5,5,15&97\%&-0.20&0.17&0.0-0.1&Coma(or Castor)   \\
   \hline
   6&38,-20,-15&4,4,12&100\%&37,-25,-12&4,4,14&98\%&-0.24&0.21&0.1-0.2& A08(new feature) \\
   \hline
   7&-57,-45,-16&5,5,13&100\%&-55,-50,-16&5,5,14&98\%&-0.22&0.29&0.2-0.3&F08(F.3 green)\\
   \hline
   8&57,-5,-10&3,3,12&93\%&60,-5,-10&5,5,18&87\% &-0.28&0.18&0.2-0.3& \\
   \hline
   9&-56,7,1&3,3,21&97\%&-50,4,3&3,3,20&90\%&-0.28&0.17&0.1-0.2& \\
   \hline
   10&30,-5,-11&3,3,6&87\%&28,0,-11&6,6,5&86\%&-0.25&0.13&0.2-0.3 &\\
   \hline
   11&&&&65,-20,-7&5,8,16&63\%&&&0.2-0.3&\\
   \hline
   12&-53,-19,-12&3,3,21&92\%&-50,-20,-18&3,3,10&93\%&-0.20&0.24&0.2-0.3&\\
   \hline
   13&&&&20,-50,-5&3,3,22&82\%&&&0.1-0.2& \\
   \hline
   14&-16,-15,-12&4,3,10&90\%&-18,-15,-16&4,3,5&91\%&-0.13&0.10&0.0-0.1&IC 2391   \\
   \hline
   15&9,-15,-9&5,6,12&97\%&&&&-0.09&0.22&0.0-0.1&  \\
   \hline
   16&42,8,-29&3,3,15&92\%&&&&-0.20&0.22&0.2-0.3& \\
   \hline
   17&&&&55,-50,10&4,4,2&86\%&&&0.2-0.3&\\
   \hline
   18&&&&20,-20,-15&5,5,4&90\%&&&0.1-0.2&A08(T3.No7) \\
   \hline
   19&&&&8,-27,-8&5,5,13,2&93\%& &&0.2-0.3&\\
\hline
   20&&&&-75,-46,-13&5,5,11&91\%& &&0.1-0.2&A08(T3.No14)\\
\hline
   21&&&&-17,-49,-10&5,5,13&82\%& &&0.2-0.3&\\
\hline
   22&&&&-69,-2,-10&5,5,22&90\%& &&0.1-0.2&A08(T3.No13)\\

  \noalign{}
  \enddata
 \end{deluxetable}

\begin{deluxetable}{ccccccccc}
 \tabletypesize{\scriptsize} \tablenum{2} \tablewidth{0pt}
  \tablecaption{The center velocity in literature corresponding groups in resent work. The column 1 is the sequence number of the groups in our work.
  Column 2 is from Dehnen(1998); Column 3 is from
  Eggen(1971, 1991, 1992a, b, c, 1996); column 4 is from Fux(2001); column 5 is from Famaey(2005); column 6 is
  according to Famaey(2007); column 7 is given by
  Famaey(2008); column 8 is from
  Klement(2008); column 9 is from Antoja (2008)
  } \tablehead{\colhead{No}&\colhead{D(U,V)}&\colhead{E(U,V)}
&\colhead{Fu(U,V)}&\colhead{F05(U,V)}&\colhead{F07(U,V)}&\colhead{F08(U,V)}&\colhead{K08(U,V)}&\colhead{A08(U,V)}}
 \startdata
  \noalign{}
   1&-40,-20&-40.4,-16.0&-42,-18&&-37,-17&-35,-18&-25,-15$^a$&Tab.3 No2 \\
   \hline
   2&-12,-22&-11.6,-20.7&-13,-19&&-15,-25&-16,-23&-25,-15$^a$&Tab.3 No1 \\
   \hline
   3&&-30.0,-50.0&-35,-45&-42,-51&-30,-50&-35,-51&-20,-50&Tab.3 No16\\
   \hline
   4&9,3&14.9,1.4&9,3&6.5,3.9&10,-5&5,1.5&6,4&Tab.3 No3 \\
   \hline
   5&-10,-5&&-3,-4&&-10,-10&&&Tab.3 No4 \\
   \hline
   6&&&&&&&&35,-20  \\
   \hline
   7&&&&&&-55,-51&&  \\
   \hline
   8&&&&&&&& \\
   \hline
   9&&&&&&&& \\
   \hline
   10&&&&&&&& \\
   \hline
   11&&&&&&&\\
   \hline
   12&&&&&&&&\\
   \hline
   13&15,-60&5.8,-59.6&&&&&&Tab.3 No6 \\
   \hline
   14&&-20.8,-15.9&&&&&&Tab.3 No12  \\
   \hline
   15&&&&&&&& \\
   \hline
   16&&&&&&&& \\
   \hline
   17&&&&&&&&\\
   \hline
   18&20,-20&&&&&&&Tab.3 No7\\
   \hline
   19&&&&&&&& \\
\hline
   20&-70,-50&&&&&&&Tab.3 No14 \\
\hline
   21&&&&&&&& \\
\hline
   22&-70,-10&&&&&&&Tab.3 No13 \\

  \noalign{}
  \enddata
  \tablenotetext{\it a}{The group 1 and 2 were unresolved by Klement(2008).}
 \end{deluxetable}

\end{document}